# EXPERIMENTING WITH THE NOVEL APPROACHES IN TEXT STEGANOGRAPHY

Shraddha Dulera[1], Devesh Jinwala[1] and Aroop Dasgupta[2]

[1] Department of Computer Engineering, S V National Institute of Technology, Surat, India
[1]d.shraddha@rocketmail.com, [2]dcjinwala@acm.org
[2] Bhaskaracharya Institute for Space Applications and Geo-Informatics, Gandhinagar, India

## ABSTRACT

*As is commonly known, the steganographic algorithms employ images, audio, video or text files as the medium to ensure hidden exchange of information between multiple contenders to protect the data from the prying eyes. However, using text as the target medium is relatively difficult as compared to the other target media, because of the lack of available redundant information in a text file. In this paper, in the backdrop of the limitations in the prevalent text based steganographic approaches, we propose simple, yet novel approaches that overcome the same. Our approaches are based on combining the random character sequence and feature coding methods to hide a character. We also analytically evaluate the approaches based on metrics viz. hiding strength, time overhead and memory overhead entailed. As compared to other methods, we believe the approaches proposed impart increased randomness and thus aid higher security at lower overhead.*

## KEYWORDS

*Steganography, Text Steganography, Feature coding*

## 1. INTRODUCTION

The term Steganography is derived from Johannes Trithemus' (1462-1516) literary work titled "Steganographia". The word steganography is of Greek origin and means concealed writing. The goal of steganography is to transmit a message through some innocuous carrier over a communication channel in order to effectively conceal the existence of the message. The carrier could be any of image, audio, video or text data. The image, audio or video data when used as a carrier, typically contains ample redundant data, so that while being used to hide the significant data, there is no apparent loss of generality [1] [2]. However, in the absence of any redundant information to be exploited to conceal the message, it is difficult to design the steganographic approaches that rely on using the text data alone for the purpose [1].

Despite the same, there have been various attempts at devising text steganographic algorithms. These attempts rely on exploiting either the syntactic or semantic characteristics or both, of the language under consideration [3] [4]. The carriers used in these approaches, typically are either the white spaces, or the punctuations or the actual words/phrases or their synonyms or other language characteristics.

However, no matter what approach is employed, it is essential that the optimum hiding is realized without sacrificing any loss in data or in their semantics, while enduring tolerable overhead. As per the observations elaborated further in section II, the existing techniques do suffer from such limitations of one kind or the other [2] [3].

Hence, in the backdrop of such observed limitations of the existing approaches, we explore and propose here, new approaches for text steganography. Our approaches are based on exploiting the features of English letters to hide the secret data e.g. using the characters having round shape or the straight vertical line as the distinctive elements or using the quadruple categorization of the English letters as the basis. We implement the algorithms proposed and empirically evaluate them as per the defined metrics to compare against the existing approaches. As per our results, the approaches proposed definitely achieve increased security at marginal increase in the overhead.

The rest of the paper is organized as follows: in section 2, we critically review the existing text steganography methods and highlight their limitations. In section 3, we propose the new approaches for text-based steganography. In section 4, first we analyze our approaches in general and then show the experimental setup used to empirically evaluate the algorithms and analyze the results, whereas discuss the probable further explorations and conclude the paper in section 5.

## 2. THEORETICAL BACKGROUND

In this section, we survey the common text steganographic approaches with an aim to critically examine them and ascertain their utility.

### 2.1. Overview

In Steganography, when aiming to hide some significant data (the subject data) in a document to protect it, it is necessary to use some other redundant data (the cover data) as a cover for the existing valid data. As mentioned before, the probable media that can be used as a cover are text, image or a movie clip, or sound bites. Obviously, higher the ratio of cover data to the subject data, easier it is to hide the subject data. Now, a text file requires lesser attributes to be stored about the data and so the eventual structure of a text file more or less resembles that of the actual data content of the same. This however, is not so, in the video, image or movie clip files as various attributes about the actual data is required to be stored. This in general makes, handling such files all the more complicated as compared to handling a text file. Hence, with relatively lesser demands on storage and bandwidth, text data must always be preferable for being used in steganography as compared to the other approaches.

However, without the need for storing detailed attributes of the underlying data, a text document lacks the necessary redundant information resulting into lower cover data to the subject data ratio, as compared to the same in an image, movie clip or an audio document. Hence, the general perception is that the text steganography is the most difficult to realize amongst all the steganographic techniques [5].

### 2.2. Related Work

In this section, we survey different approaches in the contemporary research that demonstrate the viability of text steganography.

In Open Space methods [6][4], extra white-spaces are inserted into the text, to achieve hiding. The white-spaces, that can be placed at the end of each line, at the end of each paragraph or between the words, exorbitantly increase the overhead.

In Semantic method [4], semantic characteristics of the language are exploited to impose redundancy. That is, synonyms of certain words are used to camouflage the actual data content, thereby hiding the information in the text. A major advantage of this method is the protection of information in case of retyping or using OCR programs. However, when used unscrupulously, this method may alter the meaning of the text.

The Feature Coding methods proposed in [2] [3], exploit syntactic alterations to impose the required redundancy. For example, characters such as h, d, b, are elongated or shortened a little thereby hiding information in the text. In addition, the text attributes (e.g. color) are altered to hide the characters.

Yet another syntactic approach proposed in [6] exploits the inherent ambiguity in the punctuation in English language, to achieve the hiding e.g. the phrases "bread, butter, and milk" and "bread, butter and milk" are both considered correct usage of commas in a list. The syntactic method is based on exploiting the fact that the choice of form is arbitrary. Alternation between forms can represent binary data.

In [7], secret data is hidden by exploiting the form of text document. This approach can range from simple to very complicated depending on the specifications. In the simplest form, the first letter of every word of paragraph is used to hide the information.

In [7] [3], the authors exploit the case insensitivity of the HTML tags to hide the secret information within. For example, the tags <p align="center">, <p align="cenTER">, <p align="Center"> and <p aLigN="center">, are all valid tags.

The Line Shifting method which is proposed in [8], the lines of the text are vertically shifted to some degree (for example, each line shifts 1/300 inch up or down) to hide the information by creating a unique shape of the text with the up or down shifting depending on the bit value in the payload. The odd lines are considered as control lines and used while decoding.

[5] and [9] propose a word shifting approach in which a word in the text is shifted horizontally left or right depending on the secret data. The advantage is reduced susceptibility to detection as change of distance between the words in a line is quite common.

In Abbreviation approach [4], abbreviation of words or phrases is used to hide secret data. Depending on secret bit abbreviated form or full form of words or phrases can be used to hide the secret data. In this method, very little information can be hidden in the text.

Other methods used for data hiding are Random Character and Word Sequence approach proposed in [4] [9]. To avoid the comparison with a known plain text and the subsequent known cover attacks, the authors propose generating independent cover texts. However, if the characters or words sequence is random giving rise to illegitimate text, it remains vulnerable to being noticed.

In [1], information hiding is done by exploiting different standards prevailing for spelling some words. For example, the spelling of dialog has different connotations in the form of dialogue or dialog. Hence, hiding is obtained by substituting one for the other, however without much advantage.

Alternatively, the text steganography may also exploit the characteristics of a specific language too. In Persian/Arabic Text Steganography [4], data is hidden by considering the existence of too many points in Persian and Arabic text. To hide bit 1, point of letter is shifted little upward and to hide bit 0, the point of letter remains unchanged.

In Hindi Language Text Steganography [9], secret bits are hidden by using the Hindi text. Hindi language has a combination of letters and letter diacritics. It also has compound letters. These characteristics are exploited with bit 0 encoding a vowel & consonants whereas bit 1 encoding the letter diacritics and compound letters.

However, restricting to the English language, we observe that the hiding efficiency in the simple approaches that we propose is better as compared to all of the approaches mentioned above. We describe our approaches in the next section.

## 3. THE PROPOSED APPROACHES

Upon analyzing the existing text steganographic approaches, we propose new approaches for text-based steganography. The approaches that are explained in this section are based on features of English language. When using all the approaches, we assume that the input to the proposed approach is a text file containing the secret message to be hidden and this message is converted to binary bits, before applying the method.

### 3.1. Approach based on curves in a character subheading

In this approach, we divide the English letters into two groups based on the shape i.e. whether a character has a curvature in its shape or not e.g. characters like 'B', 'C' have rounded shape whereas the letters 'A', 'E' etc. do not have so.

The entire classification based on this logic is as shown in Table 1. Using this approach; we can hide data in two ways as follows:

Table 1. Groups: Based on Round Shape/Curve

| Group ID | Group Name | Bit to be hidden | Letters used |
|---|---|---|---|
| A | Letters with full/partial curvature | 0 | B, C, D, G, J, O, P, Q, R, S, U |
| B | Letters without any sort of curvature | 1 | A, E, F, H, I, K, L, M, N, T, V, W, X, Y, Z |

#### 3.1.1 Using Random Character Sequence

We generate a random string that contains the single letters (from alphabet) as the cover text. Subsequently, whenever we want to hide a '0' bit in the input text file, we use the letters from the group A amongst the letters generated; whereas whenever we wish to hide a '1' bit, we use the letters from the group B amongst the letters generated.

#### 3.1.2 Using Sentence Case

An alternative to the approach above is to take a sentence and use the first letter from every sentence as the cover text. Subsequently, for hiding a '0' bit in the input text file, we check the first letter in the sentence. If the first letter of sentence is from first group then we can hide a '0' bit and if first letter of sentence is from other group then we can hide '1' bit. However, if it is not so, then we introduce some word in the sentence that starts with the desired letter e.g. if we want to hide the message having bits say "110". And we select a cover text sentence that is viz.

"*All birds can fly. Ostrich is a bird. Ostrich can also fly.*"

Now, to hide the bit '1', we can select the first letter of the first word i.e. 'A' which appropriately is the member of group B. However, we cannot hide the next bit which is also a '1' using the first letter of the beginning of the next sentence – which is 'O' belonging to group A. Therefore, we modify the next sentence as follows:

"*All birds can fly. This is a bird. Ostrich can also fly.*"

Now, since the first letter of the second sentence is 'T' – we can hide the bit '1' using it. Similary, for the bit '0', the letter 'O' of the next sentence can easily be used. Therefore the bits "110" would be replace with the statement as shown above.

### 3.2. Approach based on vertical straight line

In this approach the basic logic is the same as discussed before, instead of round shape or curve, we use straight vertical line in a character as the basis to group English letters.

That is, we observe that some English letters contain one vertical straight line e.g. I, J etc. whereas some others either do not contain a straight line or contain more than one single line. Grouping in this manner, we obtain the character groups as shown in Table 2.

Table 2. Groups: Based on Vertical Straight Line

| Group ID | Group Name | Bit to be hidden | Letters used |
|---|---|---|---|
| A | Number of vertical lines != 1 | 0 | A, C, G, H, M, N, O, Q, S, U, V, W, X, Y, Z |
| B | Number of vertical lines == 1 | 1 | B, D, E, F, I, J, K, L, P, R, T |

Table 3. Groups: Based on Quadruple Characterization

| Group ID | Group Name | Letters in group | Bits used |
|---|---|---|---|
| A | Curved Letters | C, D, G, O, Q, S, U | 00 |
| B | Letters with middle horizontal straight line | A, B, E, F, H, P, R | 01 |
| C | Letters with one vertical straight line | I, J, K, L, T, Y | 10 |
| D | Letters with diagonal line | M, N, V, W, X, Z | 11 |

Using this approach again, we can hide data in all the ways discussed before i.e. using the Random Character Sequence OR using the Sentence case.

### 3.3. Quadruple categorization

In this approach, we make four groups of English letters based on the features of letters. That is we group the English letters into four groups based on whether the letter has a curve, middle horizontal straight line, a single straight vertical line or multiple straight vertical lines. The groupings and the bits used for the same are as shown in the Table 3. Using this approach again, we can hide data in all the ways discussed earlier.

## 4. EVALUATION AND ANALYSIS

In all these proposed approaches, we have classified the characters in various groups using a classification based on simple shapes. Hence, an initial effort is required to categorize a character in the appropriate group e.g. whenever we wanted to hide secret bits using the random character sequence method say, we search for particular letter based on the value of the secret bit. If a letter is not from that particular group, then we have to skip that letter and consider next letter to hide bit. Hence, the time complexity of all the proposed approaches is $O(1)$ to hide a bit if we select the letter according to the secret bit. On the same lines, the worst case time complexity is $O(N)$ for all proposed approaches to hide a bit.

In addition, it is emphasized that though using random character sequence to hide the secret bits, the encoded text may become noticeable, but at the same time, because of the randomness, decoding it is made non-trivial. Hence, the security of the schemes is largely derived from the randomness imparted.

Now, we discuss the methodology of evaluation, parameters for comparison, platform and tools used for the empirical evaluation of the proposed approaches. We also compare the empirical results we obtained with those prevailing in the literature – specifically with those in [4].

One of the vital parameter for evaluating the text steganographic algorithms is the *capacity* – that is the amount of data bits that can be hidden in the cover text. Hence, we measure the number of bytes hidden by the existing methods and by all the approaches we propose, for the specific cover text. *Time* and *memory overhead* are other two important factors, for comparison. Hence, we have checked the memory overhead and time overhead needed to execute these methods and compared them. We have taken five different secret message size and cover text size and we have calculated time overhead and memory overhead of various approaches.

In the feature coding method, we changed case of every word to proper case or lower case depending on the secret bit. In random character sequence, we have generated random character sequence and from which we chose two characters to hide '0' bit and '1' bit.

For the implementation, we have used random character sequence as a cover text. We have generated random character sequence of particular length and we are hiding secret bits in this generated random character sequence. We have used Microsoft visual studio for implementation and Dottrace 3.1 profiler for memory profiling and time profiling. Tables 4 to 8 show the results of our experimentation.

Based on these observation results, we show a graphical comparative view in the graphs in figures 1 to 5.

In Fig. 1 we depict, the number of bytes hidden by the existing methods as well as by the proposed approaches using random character sequence. As can be observed, using the approach which is based on curves in a character subheading we can hide a higher number of bytes. On the other hand, least number of bytes can be hidden by Inter-sentence space approach because space between two sentences is used to hide secret bits.

We show the comparative outlook at the number of bytes hidden using various approaches, with a message text size of 1000 bytes. It can be seen that the number of bytes hidden using feature coding is more as compared to the other existing methods. Also, the number of bytes hidden using the approach based on curves in a character subheading is higher as compared to all the other implemented methods.

In Figure 3, we show the maximum cover text size required when message text size is 200 bytes. As can be seen, the Inter-Sentence space method requires maximum cover text size to hide 200 bytes, whereas minimum cover text size is required to hide 200 bytes using the approach based on curves in a character subheading.

We show the time overhead and memory overhead of some of the existing and all of the proposed approaches using random character sequence. As observed earlier, among all the existing methods which we have implemented, we can hide more number of bytes using feature coding method. So we consider this method as the base method and compare the time overhead and memory overhead our proposed approaches with this method.

We depict the additional time and memory overhead of the proposed approaches as compared to the feature coding method, when the message size is 1000 bytes. As can be observed, the number of bytes hidden by the proposed approaches is higher, whereas the time and memory overhead of the proposed approaches is marginally less compared to feature coding approach.

Lastly, we note that the security of the open space methods (Inter-sentence space and Inter-word space) and feature coding method is as such poor as the approach used therein is noticeable. However, the security of random character sequence and all other proposed approaches is high because they are based on randomness.

## 5. CONCLUSION

In this paper, we propose newer approaches for text-based steganography for English language texts. In these approaches, we exploit the shapes of the English characters to hide secret bits. We can hide more number of bytes using proposed approaches, while additional time overhead and memory overhead is also minimal compared to feature coding method. These approaches are applicable to the soft-copy texts as well as hard-copy texts. Our analysis reveals that our approaches impart increased randomness and because of randomness, these approaches are noticeable but it cannot be decoded until a user is not aware about these approaches. In addition, the proposed approaches are also immune to retyping and reformatting of text.

However, one of the weaknesses of the proposed approaches is that once their applicability is known, they can easily be attacked. Hence, it is essential to keep the application of a particular approach to a particular data set secret, while using them.

Because of randomness, the use of random character sequence to hide secret data is noticeable. So we have also mentioned other ways to hide secret data using sentences. When we have used sentence case, we have skipped sentences if sentences are not according to secret data. Yet another way to hide secret data is incorrect grammar approach. In this approach, we have randomly placed word as starting word of sentence. We have randomly selected words depend on secret bits and that is why sentence can be meaningless sometime. So both these ways are again noticeable. Using these proposed approaches manually turns out to be too tedious if we take care about syntax and that is we resort to linguistic steganography in which the syntax of a sentence is taken care of. In future, we intend to carry out the formal security analysis of the methods proposed as well as to extend this work to linguistic steganography in which the syntax of a sentence and sequence of sentence can also be taken care of, so that the sentence and the paragraph, both will be true grammatically.

## ACKNOWLEDGEMENTS

The authors would like to thank all the anonymous reviewers who helped refine the state of this paper. The author would also like to acknowledge the Department of Information Technology, Ministry of Information and Communication Technology, Govt of India for funding this research publication.

# REFERENCES

[1] M.H. Shirali-Shahreza, "Text steganography by changing words spelling," Proc. 10th Int. Conf. on Advanced communication technology, 2008, pp. 1912-1913.

[2] M. Shirali-Shahreza, M. H. Shirali-Shahreza, "Text steganography in SMS," Proc. Int. Conf. Convergence Information Technology, Washington, 2007, pp. 2260-2265.

[3] K.F. Rafat, "Enhanced text steganography in SMS," Proc. of the 2nd Int. Conf. Computer, Control and Communication, Karachi, 2009, pp.1-6.

[4] M. H. Shirali-Shahreza, M. Shirali-Shahreza, "A new approach to persian/arabic text steganography," Proc. 5th Int. Conf. Computer and Information Science, Washington, 2006, pp. 310-315.

[5] J. T. Brassil, S. Low, L. O'Gorman, and N. F. Maxemchuk, "Electronic marking and identication techniques to discourage document copying," IEEE Journal on Selected Areas in Communications, vol. 13, no.8, 1995, pp. 1278-1287.

[6] W. Bender, D. Gruhl, N.Morimoto, and A Lu, "Techniques for data hiding," IBM Systems Journal, vol. 35, nos. 3&4, 1996, pp. 313-336.

[7] A. Gutub, M. M. Fattani, "A Novel Arabic Text Steganography Method Using Points and Extensions," Proc. of the WASET, 2007, pp. 28-31.

[8] L. Robert, T.Shanmugapriya, "A study on digital watermarking techniques," Int. Journal of Recent Trends in Engineering, vol.1, no.2, May 2009.

[9] K. Alla and R.S.R. Prasad., "An evolution of hindi text steganography," Proc. of the 6th Int. Conf. Information Technology: New Generations, Washington, 2009, pp. 1577-1578.

Table 4. Overload in various text steganography approaches: Text size in 200 Bytes

| Text Steganography Approach | Message Text size (Bytes) | Cover Text size (Bytes) | No. of bytes can hide (Bytes) | Time Overhead (ms) | Memory Overhead (ms) |
|---|---|---|---|---|---|
|  | - | - | 0 | - | - |
| Feature Coding | 200 | 660 | 13 | 18,158 | 724,846 |
| Inter Sentence space | 200 | 660 | 1 | 19,276 | 731,860 |
| Inter Word space | 200 | 660 | 14 | 20,906 | 730,850 |
| Random Character Sequence | 200 | 660 | 14 | 28,100 | 756,110 |
| Method based on round shape | 200 | 660 | 43 | 22,369 | 780,206 |
| Method based on vertical straight line | 200 | 660 | 41 | 22,480 | 802,600 |
| Quadruple Categorization | 200 | 660 | 34 | 27,438 | 838,056 |

Table 5. Overload in various text steganography approaches: Text size 400 Bytes

| Text Steganography Approach | Message Text size (Bytes) | Cover Text size (Bytes) | No. of bytes can hide (Bytes) | Time Overhead (ms) | Memory Overhead (ms) |
|---|---|---|---|---|---|
|  | - | - | 0 | - | - |
| Feature Coding | 400 | 1320 | 33 | 21,392 | 865,934 |
| Inter Sentence space | 400 | 1320 | 2 | 20,598 | 726,961 |
| Inter Word space | 400 | 1320 | 29 | 21,833 | 899,150 |
| Random Character Sequence | 400 | 1320 | 31 | 32,088 | 904,374 |
| Method based on round shape | 400 | 1320 | 86 | 23,377 | 911,872 |
| Method based on vertical straight line | 400 | 1320 | 82 | 23,266 | 915,474 |
| Quadruple Categorization | 400 | 1320 | 77 | 24,700 | 903,730 |

Table 6. Overload in various text steganography approaches: text size 600 bytes

| Text Steganography Approach | Message Text size (Bytes) | Cover Text size (Bytes) | No. of bytes can hide (Bytes) | Time Overhead (ms) | Memory Overhead (ms) |
|---|---|---|---|---|---|
|  | - | - | 0 | - | - |
| Feature Coding | 600 | 1980 | 52 | 20,289 | 724,624 |
| Inter Sentence space | 600 | 1980 | 3 | 17,404 | 739,728 |
| Inter Word space | 600 | 1980 | 45 | 22,604 | 773,498 |
| Random Character Sequence | 600 | 1980 | 41 | 33,521 | 828,968 |
| Method based on round shape | 600 | 1980 | 126 | 24,810 | 862,116 |
| Method based on vertical straight line | 600 | 1980 | 119 | 24,920 | 909,768 |
| Quadruple Categorization | 600 | 1980 | 110 | 24,369 | 943,780 |

Table 7. Overload in various steganography approaches: text size 800 bytes

| Text Steganography Approach | Message Text size (Bytes) | Cover Text size (Bytes) | No. of bytes can hide (Bytes) | Time Overhead (ms) | Memory Overhead (ms) |
|---|---|---|---|---|---|
|  | - | - | 0 | - | - |
| Feature Coding | 800 | 2640 | 66 | 18,180 | 938,843 |
| Inter Sentence space | 800 | 2640 | 4 | 17,294 | 732,529 |
| Inter Word space | 800 | 2640 | 58 | 20,825 | 944,614 |
| Random Character Sequence | 800 | 2640 | 48 | 31,292 | 931,979 |
| Method based on round shape | 800 | 2640 | 172 | 37,996 | 748,077 |
| Method based on vertical straight line | 800 | 2640 | 161 | 27,533 | 776,356 |
| Quadruple Categorization | 800 | 2640 | 146 | 26,542 | 826,065 |

Table 8: Overload in various text steganography approaches: text size 1000 bytes

| Text Steganography Approach | Message Text size (Bytes) | Cover Text size (Bytes) | No. of bytes can hide (Bytes) | Time Overhead (ms) | Memory Overhead (ms) |
|---|---|---|---|---|---|
| | - | - | 0 | - | - |
| Feature Coding | 1000 | 3564 | 90 | 17,850 | 738,458 |
| Inter Sentence space | 1000 | 3564 | 6 | 17,404 | 746,610 |
| Inter Word space | 1000 | 3564 | 79 | 21,926 | 785,326 |
| Random Character Sequence | 1000 | 3564 | 76 | 32,504 | 841,634 |
| Method based on round shape | 1000 | 3564 | 232 | 32,599 | 754,809 |
| Method based on vertical straight line | 1000 | 3564 | 220 | 30,617 | 791,462 |
| Quadruple Categorization | 1000 | 3564 | 205 | 32,269 | 833,271 |

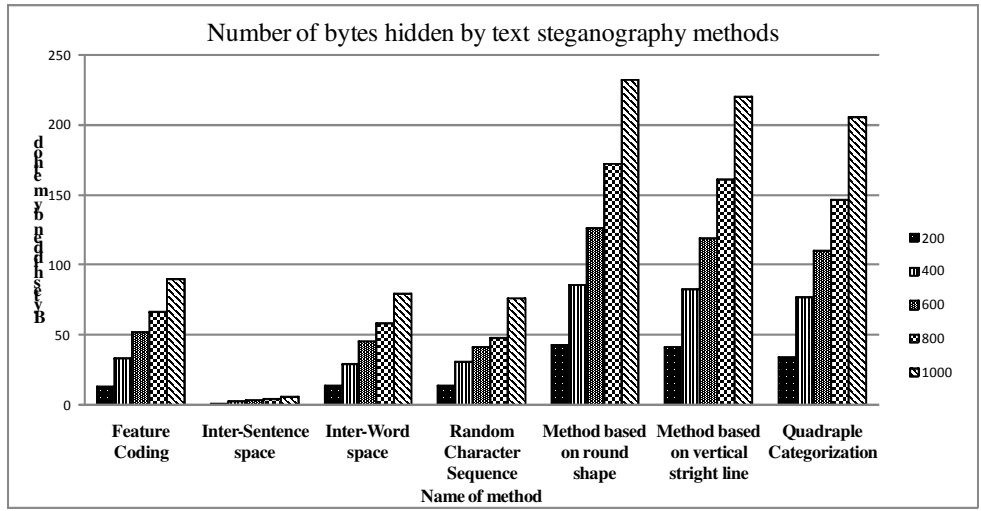

Figure 1. Number of bytes hidden using various text steganography methods

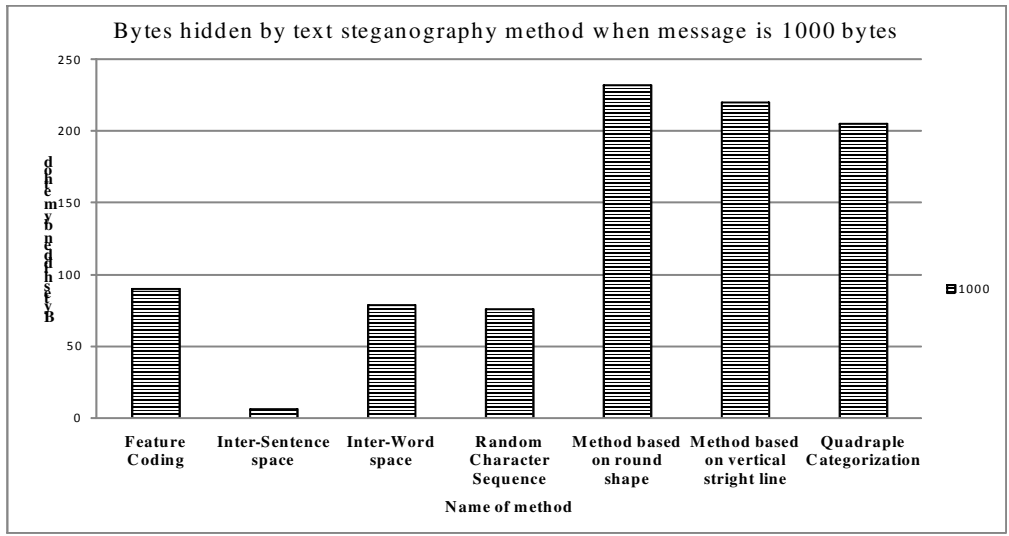

Figure 2. Number of bytes hidden when message size is 1000 bytes

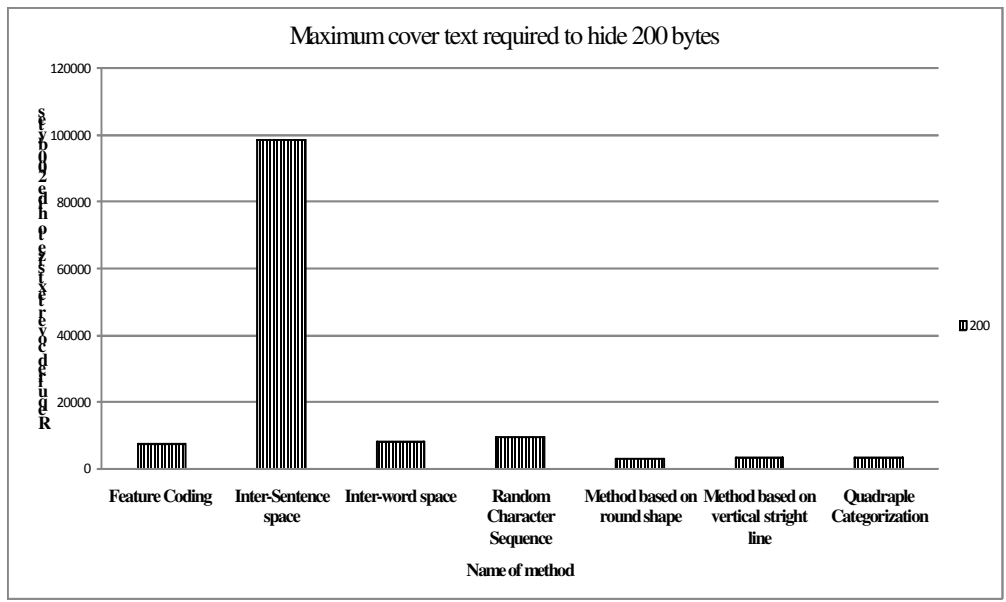

Figure 3. Maximum cover text required to hide 200 bytes

International Journal of Network Security & Its Applications (IJNSA), Vol.3, No.6, November 2011

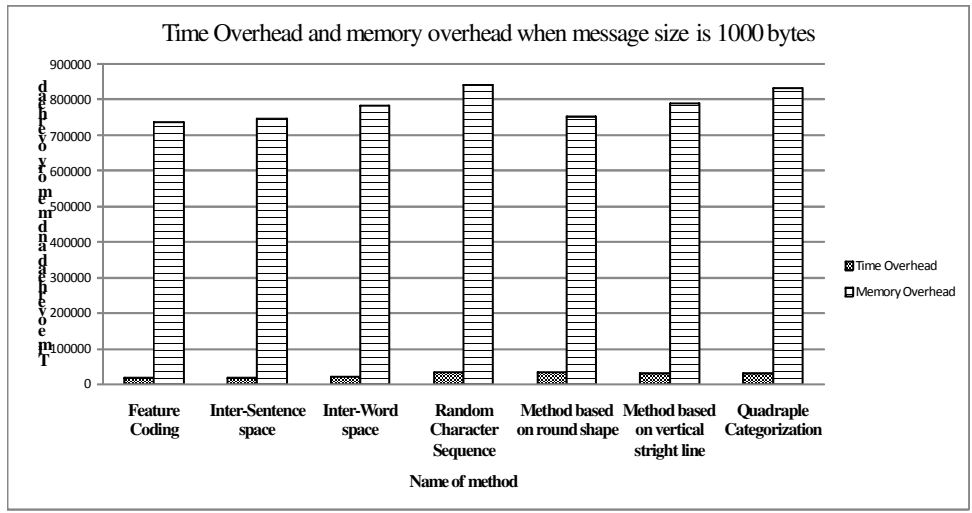

Figure 4. Time overhead and Memory overhead when message size is 1000 bytes

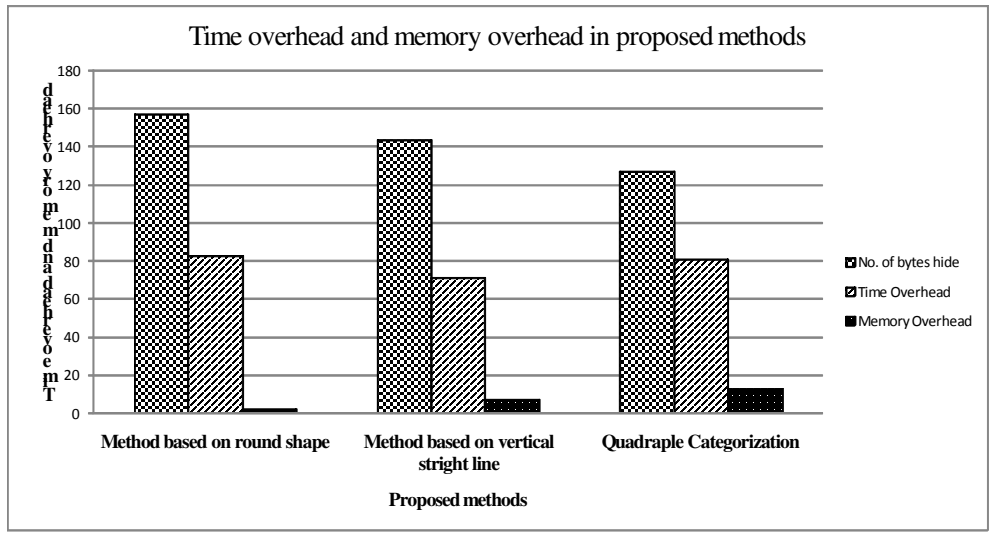

Figure 5. Additional time and memory overhead in proposed approaches

**Authors**

Shraddha Dulera is working as Senior Research Fellow at Dhirubhai Ambani Institute of Information and Communication Technology. She has received the Master of Technology in Computer Engineering from National Institute of Technology, Surat. Her main research areas are Information Security and Biometrics.

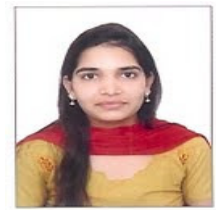

Devesh Jinwala is an Associate Professor in Computer Engineering Department at S V National Institute of Technology (SVNIT), Surat. He holds a PhD in Computer Engineering. His research interests lie in

Information Security and Privacy, Use of Ontologies in Software Engineering and Resource Discovery & Load Balancing in Distributed Systems. He has more than 50 publications in Proceedings of the referred International Conferences & Journals. He also has two ongoing research projects funded by the Indian Space Research Organization and Department of Information Technology, Ministry of Information and Communication Technology, Govt of India. He is currently the Head of the Department of Computer Engineering at SVNIT.

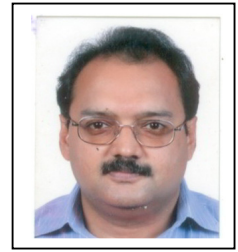

Arup Ranjan Dasgupta received his Post Graduate Degree in Electrical Communication Engineering in 1970 from the Indian Institute of Science, Bangalore. He worked in the Satellite Instructional Television Experiment (SITE) of Space Applications Centre, Indian Space Research Organization, contributing to the design of rugged television receivers installed in rural and remote areas under the SITE. Mr Dasagupta is also Associate Programme Director for the GRAMSAT that involves the use of satellite communication technologies for distance learning, telemedicine and information networking. He has operationalised the INSAT mobile satellite service system. Mr Dasagupta has nearly 50 technical publications to his credit. Mr Arup Ranjan Dasagupta is presented the ASI Award for the year 2000 in recognition of his significant contribution to Space Science and Applications.

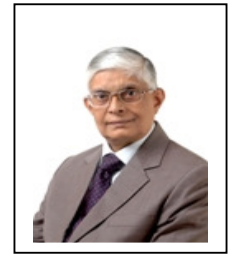